 \documentstyle[prl,epsf,aps]{revtex}

\begin{document}

\draft % To print out PACS

%%%
 \twocolumn[{
%%%
 \widetext

\title{Electron-lattice relaxation, and soliton structures and their interactions in 
polyenes.}

\author{Robert J. Bursill$^1$ and William Barford$^2$
}

\address{
$^1$School of Physics, University of New South Wales, Sydney, NSW 2052,
Australia.
$^2$Department of Physics and Astronomy, The University of Sheffield, Sheffield, 
S3 7RH, United Kingdom.
}

\maketitle

%%%
 \mediumtext

\begin{abstract}
Density matrix renormalisation group calculations of a
suitably parametrised
 model of long polyenes (polyacetylene oligomers), which incorporates both long
 range Coulomb interactions and adiabatic lattice relaxation, are
 presented.  The $1^3B_u^+$ and $2^1A_g^+$ states are found to have a
2-soliton and 4-soliton form, respectively, both with large relaxation energies. The 
$1^1B_u^-$ state
 forms an exciton-polaron and 
has a very small relaxation energy. The relaxed energy of the
$2^1A_g^+$ state lies below that of the $1^1B_u^-$ state.
The soliton/anti-soliton
pairs are bound.
\end{abstract}


%%%
 }]
%%%
 \narrowtext

Electronic interactions in polyenes and polyacetylene (PA) induce strong 
spin-density-wave correlations in the ground state, resulting in low energy spin-flip 
(or covalent) triplet ($^3B_u^+$) excitations. These combine to form even-parity 
(dipole-forbidden) singlet ($^1A_g^+$) excitations. Optical (dipole-allowed) 
transitions to the odd-parity singlet state ($^1B_u^-$) are essentially ionic in 
character, resulting in charge transfer from one site to another. In the non-interacting 
limit the $1^3B_u^+$ and $1^1B_u^-$ states are degenerate, and the $2^1A_g^+$ 
state always lies higher in energy. However, electron 
correlations can lead to a reversal of the energetic ordering of the $1^1B_u^-$ and  $2^1A_g^+$ 
states. Electron-electron correlations in 
$\pi$ conjugated systems, such as PA, are conveniently modelled by the one-band 
Pariser-Parr-Pople (P-P-P) model, which includes long range Coulomb interactions.

Electron-phonon interactions in the non-interacting limit are described by the SSH 
model. In the adiabatic limit it predicts a wealth of 
non-linear excitations, including charged/spinless ($S^{\pm}$) and neutral/spin 1/2 
($S^{\sigma}$) solitons. It is the inter-play of both electron-electron and electron-phonon interactions in PA which leads to an extremely rich variety of excitations. To describe these excitations we employ the density matrix renormalisation group (DMRG) \cite{White} 
method to solve the P-P-P-SSH model, and utilise the Hellmann-Feynman  
(H-F) theorem to calculate the low-lying excited states and the lattice relaxation 
associated with them. 

Earlier work on the solitonic structure of the low-lying excitations include, a 
renormalisation group calculation of the Hubbard-SSH model of up to 16 sites 
\cite{hayden86}; a mean-field study of the Heisenberg-Peierls model 
\cite{takimoto89}; an exact diagonalisation of a 12 site extended Hubbard-SSH 
model \cite{gammel93}; and a strong coupling and perturbation calculation of the 
Hubbard-SSH model \cite{su95}. The DMRG method has recently been used by 
and Yaron et al.\ \cite{yaron98} and Fano et al.\ \cite{fano98} to solve the P-P-P 
model for linear and cyclic polyenes, respectively. Jeckelmann \cite{jeckelmann98} 
studied the metal-insulator transition in doped PA by solving the extended Hubbard-SSH with the DMRG method. Likewise, Kuwabara et al.\ \cite{kuwabara98} used 
the DMRG method to study the relative stability of bipolarons using the same model.

The P-P-P-SHH Hamiltonian is defined as
\begin{eqnarray}
{\cal H}
& = &
- 2\sum_{i=1}^{N-1} t_i \hat{T}_i
+ \frac{1}{4 \pi t_0 \lambda} \sum_{i=1}^{N-1} \Delta_i^2 + 
\Gamma \sum_{i=1}^{N-1} \Delta_i \nonumber \\
& &
+\;
U \sum_{i=1}^N \left( n_{i\uparrow}- 1/2 \right)
\left( n_{i\downarrow} - 1/2 \right)
%U \sum_{i=1}^N n_{i\uparrow} n_{i\downarrow}
\nonumber
\\
& &
+\; \frac{1}{2} \sum_{i \ne j}^N  V_{ij} (n_i - 1)(n_j - 1),
\end{eqnarray}
where,
$t_i = \left( t_0 + \frac{\Delta_i}{2} \right) $ and $\hat{T}_i
= \frac{1} {2} \sum_{\sigma}
(c_{i+1 \sigma}^{\dagger} c_{i \sigma} +  h.c.)$
is the bond order operator of the $i$th.\ bond. We use the Ohno function for the 
Coulomb interaction:
$V_{ij} = U / \sqrt{ 1 + \beta r_{ij}^2 }$,
where $\beta = (U/14.397)^2$ and bond lengths are in \AA. The single and double 
bond lengths used in the evaluation of $V_{ij}$ are 1.46 \AA ~and 1.35 \AA, 
respectively, and the bond angle is $120^0$. Various
semi-empirical parametrisations exist for $t_0$ and $U$.  We adopt the values which 
are optimal for benzene \cite{bursill98}, whose C-C bond length of 1.40 \AA ~is 
almost the same as the average bond length in PA thin films, i.\ e., $t_0 = 2.539$ eV 
and $U=10.06$ eV.  The dimensionless electron-phonon coupling constant, 
$\lambda$, is defined by
$\lambda =2 \alpha^2/\pi K t_0$,
where $K$ is the elastic spring constant (estimated to be 46 eV 
$\AA^{-2}$  \cite{ehrenfreund87}), and $\alpha$ relates 
the actual distortion of the $i$th.\ bond from equilibrium, $y_i$, 
to $\Delta_i$:
$ y_i = \Delta_i / 2 \alpha $.
$\Gamma$ is chosen so that the relaxed ground state of an infinite polymer has the 
same chain length as the unrelaxed state, i.\ e.,
$ \sum_{i=1}^{N-1} \Delta_i = 0 $.
This ensures that the average hopping integral is $t_0$, which is applicable to C-C 
bond lengths of 1.40 \AA. However, the chain length is permitted to change for 
excited states, and for all the states of finite oligomers. The remaining parameter, 
$\lambda$, is chosen so that the model fits the {\em vertical} excitation energies of the 
$1^1B_u^-$ and $2^1A_g^+$ states of hexatriene in the gas phase \cite{flicker}. A 
choice of $\lambda = 0.115$ and $\Gamma = 0.602$ gives 4.965 eV and 5.212 eV, 
compared to the experimental values of 4.96 eV and 5.21 eV, for the $1^1B_u^-$ and 
$2^1A_g^+$ states, respectively.

The equilibrium values of the bond length distortion are determined by the H-F 
condition that,
\begin{eqnarray}
\frac{ \partial E(\{ \Delta_i \})} {\partial \Delta_i} = 0
\Rightarrow
\Delta_i = 2\pi t_0 \lambda \left[ \langle 
\hat{T}_i \rangle - \Gamma \right].
\label{HF_condition}
\end{eqnarray}

${\cal H}$ possesses spatial reflection, particle-hole and spin-flip symmetries. 
Symmetrised eigenstates of ${\cal H}$ are constructed by an efficient process which 
makes use of the fact that the block symmetry operators commute with the density 
matrix at all stages of the calculation and which has been tested by making 
comparisons with exaxt results \cite{boman1,boman2}.

The calculation of the relaxed energy of a given state for a given 
chain length is as follows:  
(1) The eigenstate is calculated for an initial choice of $\{ t_i \}$ by building up the 
lattice to the target chain size using the infinite lattice algorithm of the DMRG 
method.  
(2)
At the target chain size the H-F condition (\ref{HF_condition}) is repeatedly applied 
until the $\{ t_i \}$ have converged.
(3)
Using the new values of $\{ t_i \}$, steps 1 and 2 are repeated. 
The procedure is succesfully terminated when the energies have converged after 
successive lattice and H-F iterations. It is necessary to sweep through the lattice after 
each set of H-F iterations to ensure that the electronic states and the lattice geometry 
have converged simultaneously.

The accuracy of the DMRG implementation has been checked in a number of ways. 
First, the method has been compared with exact results in the non-interacting $(U = 0)$ 
limit. The convergence of the ground state energy with superblock Hilbert space size 
(SBHSS) is shown in Table \ref{U=0_convergence} for the $N = 102$ site system 
with various lattice geometries. The total energy converges to within 0.005 eV which 
is sufficient for energy gaps, which are of the order of 1 eV, to be resolved to within 
1\% or better. This represents the DMRG at its 
least accurate,
as the addition of correlations improves convergence, as can be seen in Table 
\ref{DMRG_convergence}, where we present the DMRG convergence for the ground 
state energy and a number of energy gaps.

\begin{figure}[p]
\centerline{\epsfxsize=8.4cm \epsfbox{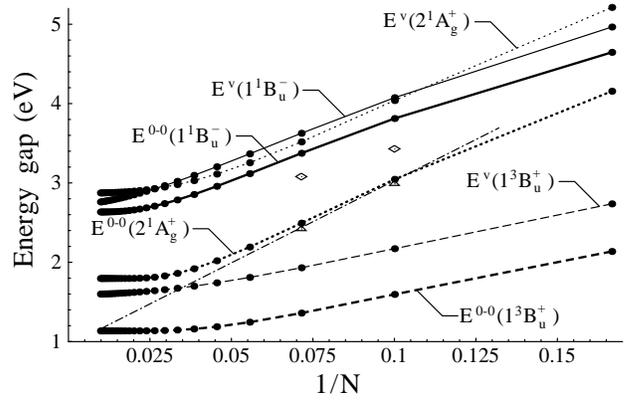}}
\caption{
Energy gaps for the 
$1^1B_u^+$ (solid lines)
$2^1A_g^+$  (dotted lines) and 
$1^3B_u^+$  (dashed lines) states as a function of $1/N$. Vertical/0-0 transitions are 
indicated by thin/thick lines.
Experimental 0-0 energies of the $1^1B_u^-$ (diamonds) and $2^1A_g^+$ (triangles) 
states for polyenes in hydrocarbon solution \protect\cite{kohler}.
The empirical fitting form $E^{\text{0-0}}(2^1A_g^+) = 0.96 + 20.72 / N$,
derived in \protect\cite{kohler} from the 8--14-site oligomer
data is also plotted (dot-dashed line).
%Also shown is the $2^1A_g^+$ energy using the relaxed $1^1B_u^-$
%geometry (thin dot-dashed line).
%
}
\label{energies}
\end{figure}

% The calculated values of $\{t_i \}$ imply that the bond length
% alternation of the ground state in the middle of the chain is 0.047 \AA.
%, close to the accepted experimental value of 0.045 \AA.
Using the ground state geometry, the vertical 
energies ($E^{\text v}$) of the 
%lowest triplet 
$1^3B_u^+$, 
%optically allowed singlet 
$1^1B_u^-$ and
%two-photon allowed singlet 
$2^1A_g^+$
 states are calculated. These, as well as the relaxed energies ($E^{\text{0-0}}$), are shown 
in Fig.\ \ref{energies} as functions of $1/N$. We first note that the vertical energies 
of the $1^1B_u^-$ and $2^1A_g^+$ states are very close, with a crossing at short 
chains, and again for long chains. In the thermodynamic limit $E^{\text v}(1^1B_u^-) 
< E^{\text v}(2^1A_g^+)$. This large $N$ crossing has also been observed in the $U$-$V$ dimerised Hubbard model \cite{boman2,shuai}.

The relaxation energy of the $1^1B_u^-$ state is modest (ca.\ 0.3 eV) and has not 
converged (i.\ e.\ it is still rapidly decreasing) for N = 102. By contrast, the relaxation 
energies of the $1^3B_u^+$ and $2^1A_g^+$ states are substantial, being ca.\ $0.5$ 
eV and $1.0$ eV, respectively, and converge rapidly with $N$. We have 
also calculated the energy of the $2^1A_g^+$ state using the relaxed geometry of the 
$1^1B_u^-$ state. This always lies lower than
$E^{\text{0-0}}(1^1B_u^-)$, which implies that a vertical
photo-excitation to the 
$1^1B_u^-$ state will decay to the $2^1A_g^+$ state. Finally, the experimental 
values of $E^{\text{0-0}}(1^1B_u^-)$ and $E^{\text{0-0}}(2^1A_g^+)$ for $N=10$ and $14$ are 
shown \cite{kohler}. The $2^1A_g^+$ values are in good agreement with our
calculation. The 
$1^1B_u^-$ values are ca.\ $0.3$ eV lower than our predictions. The
experimental results for $N = 8$--14 have been analysed by Kohler
\cite{kohler}. For the $2^1A_g^+$ state the empirical relation
$E^{\text{0-0}}(2^1A_g^+) = 0.96 + 20.72 / N$ was derived, in good
agreement with the photoinduced absorption result ca.\ 1.1 eV
for polyacetylene thin films. However, Fig.\ 1 suggests
that an algebraic fitting form is incorrect---the true scaling behaviour
is exponential, and can only be seen by considering sufficiently
large systems. 

\begin{figure}[p]
\centerline{\epsfxsize=8.4cm \epsfbox{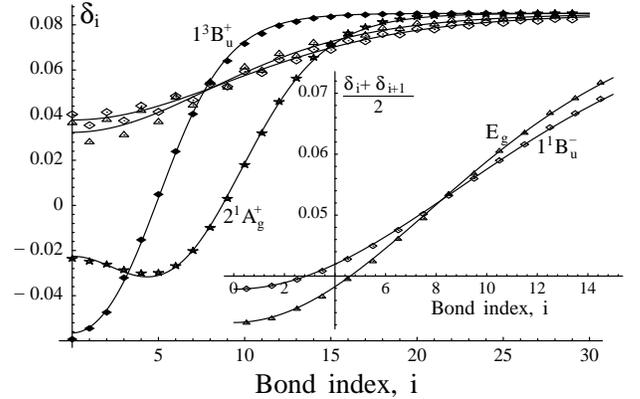}}
\caption{
The geometries ($\delta_i$ as a function of bond index $i$ from the 
center of the lattice) of various excitations: $1^1B_u^-$ (open diamonds), 
$1^3B_u^+$ (filled diamonds), $2^1A_g^+$ (stars) and the charged state, $E_g$,
(open triangles), 
for the $N = 102$ site system. The solid lines are fits to the 2-soliton form 
(\protect\ref{2_soliton_form}) (and the
4-soliton form (\protect\ref{4_soliton_form}) for the $2^1A_g^+$). The inset shows 
the two-point averages
$\left( (2i+1)/2, (\delta_i + \delta_{i+1})/2 \right)$
for the polaronic $E_g$ and $1^1B_u^-$ states,
which are well described by the 2-soliton fits.
}
\label{geometries}
\end{figure}

In Fig.\ \ref{geometries} we plot as a function of bond index from the center of the 
chain, the normalised staggered bond dimerisation, defined as,
$
\delta_i \equiv (-1)^i (t_i - \bar{t}) / \bar{t},
$
where $\bar{t}$ is the average value of $t_i$ in the middle of the chain \cite{comment0}. Note that the 
$1^3B_u^+$ and $2^1A_g^+$ states undergo considerable bond distortion, whereas 
the $1^1B_u^-$ state and the charged state (denoted $E_g$) show a weak polaronic 
distortion of the lattice. The oscillatory behaviour of $\delta_i$ in the polaronic 
distortions indicates a local expansion of the lattice. 

We fit the $1^3B_u^+$, $1^1B_u^-$ and charged state to a 2-soliton form 
\cite{su95,campbell},
\begin{eqnarray}
\delta_i
& = &
{\bar \delta} \left\{ 1 + \tanh\left( 2x_0/\xi\right)
\left[ \tanh \left( (i-x_0) /\xi \right) \right. \right.
\nonumber
\\
& &
-\; \left. \left. \tanh\left( (i+x_0)/\xi \right) \right] \right\}.
\label{2_soliton_form}
\end{eqnarray}
The $2^1A_g^+$ state, however, evidently requires a 4-soliton \cite{su95,campbell} 
fit of the form,
\begin{eqnarray}
\delta_i & = & {\bar \delta} \left\{ 1 + \tanh\left( 2x_0/\xi \right)
\left[ \tanh\left( (i-x_d-x_0)/\xi \right) \right. \right.
\nonumber
\\
& & -\;
\tanh\left( (i-x_d+x_0)/\xi\right)
+ \tanh\left( (i+x_d-x_0)/\xi\right)
\nonumber
\\
& &
-\; \left. \left. \tanh \left( (i+x_d+x_0)/\xi \right)
 \right] \right\}.
\label{4_soliton_form}
\end{eqnarray}
These functions give good fits to the relaxed geometries of the 
$N = 102$ site system, as shown in Fig.\ \ref{geometries}.
The difference in energy between using 
the fits and the actual relaxed geometry is around 0.01 eV. 
%The parameters are listed 
%in Table \ref{fitting_parameters}. 
The 4-soliton character of 
$2^1A_g^+$ state indicates the strong inter-play between electron-lattice relaxation 
and electron-electron correlations in polyenes, for, as indicated earlier, this state has a 
considerable triplet-triplet character.

\begin{figure}[p]
\centerline{\epsfxsize=8.4cm \epsfbox{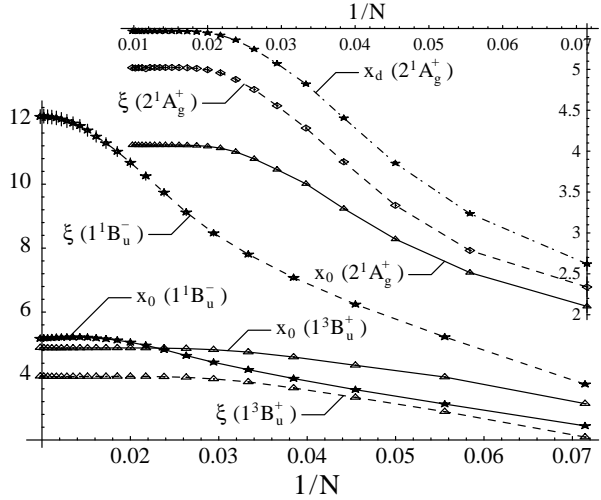}}
\caption{
The convergence of the soliton 
fitting parameters $x_0$ and $\xi$ for the $1^1B_u^-$  and $1^3B_u^+$ states
with the lattice size, $N$.  The inset shows $x_0$,  $\xi$ and $x_d$ for the $2^1A_g^+$ state.
}
\label{geometry_convergence}
\end{figure}

Fig.\ \ref{geometry_convergence} depicts the convergence of the various fitting 
parameters as a function of $N$. The $1^3B_u^+$ and $2^1A_g^+$ states converge 
rapidly with $N$, whereas the $1^1B_u^-$ state shows strong finite-size effects, and 
the coherence length $\xi$ only begins to converge at around $N = 102$.
The fact that the soliton structures converge with $N$ leads to two important 
observations: First, the soliton/anti-soliton pair are bound, because if they were not 
their separation $x_0$ would increase with $N$. 
Second, the soliton structures
 are pinned in the middle of the lattice. This is a 
consequence of the classical adiabatic treatment of the lattice, and is one
 of the 
reasons why the energy curves flatten off rapidly as $N \rightarrow \infty$.

\begin{figure}[p]
\centerline{\epsfxsize=8.4cm \epsfbox{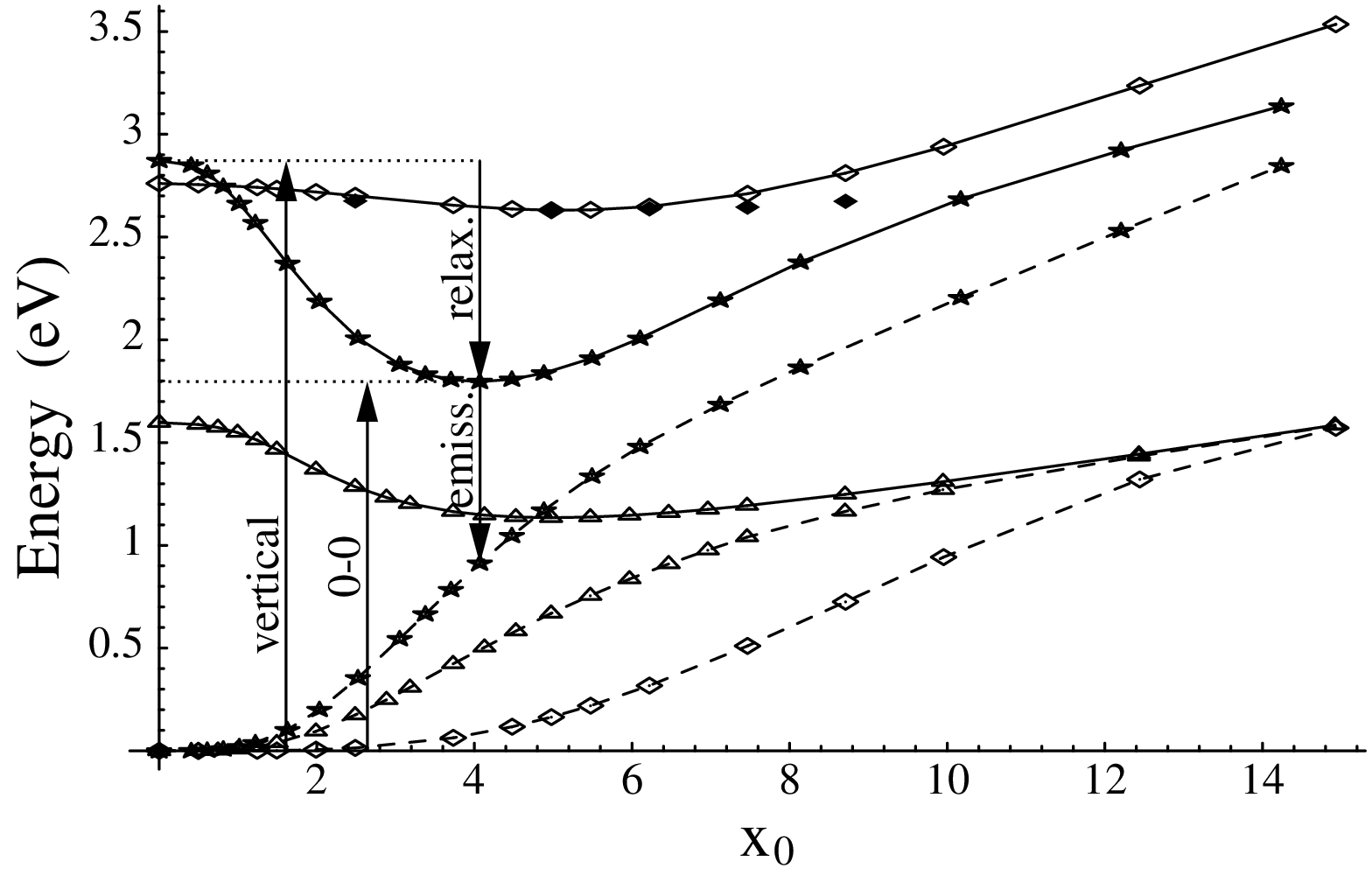}}
\caption{
Potential energy curves (solid lines) for the $1^1B_u^-$ (diamonds),
$2^1A_g^+$ (stars) and $1^3B_u^+$ (triangles) 
states for the $N=102$ site lattice. The dashed curves are the corresponding ground 
state ($1^1A_g^+$) potential energies. For the $1^1B_u^-$ and $1^3B_u^+$ cases, 
the curves are generated using the soliton pair form (\protect\ref{2_soliton_form}) 
with the fitted values of $\xi$ (4.01 and 12.12, respectively) and varying $x_0$. For 
the $2^1A_g^+$ case the curves are generated using the 4-soliton form 
(\protect\ref{4_soliton_form}) with the fitted value $\xi = 4.93$ and varying $x_0$. 
$x_d$ is chosen so that the ratio $x_d / x_0$ remains fixed at its fitted value of 
$1.35$. 
The solid diamonds are the values of the $1^1B_u^-$ energy when $\xi$ is also allowed 
to vary. 
The energies of the vertical, 0-0 and emission transitions, and the relaxation 
energy can be read off from this plot. This is illustrated using arrowed vertical 
lines for the $2^1A_g^+$ state. 
%For the $1^3B_u^+$ state the energy reduction from varying $\xi$ is negligible
% over this range of $x_0$ values.
%
}
\label{potential_energy_curves}
\end{figure}

To further investigate the soliton/anti-soliton interactions, adiabatic
 potential energy curves \cite{su95} (i.\ e.\ the energy as a function
 of soliton separation, $x_0$) are plotted in
 Fig.\ \ref{potential_energy_curves}. Our results differ qualitatively from
 previous approximate calculations \cite{su95} in that the $1^1B_u^-$ and
 $1^3B_u^+$ are bound---the potentials have a minimum and are attractive
 for large $x_0$. This attractive soliton/anti-soliton interaction implies much
 stronger binding for the $2^1A_g^+$ state than that obtained in
 \cite{su95}, where the binding energy was found to be
 ca.\ 0.05 eV.
It should be noted, however, that \cite{su95} uses a Hamiltonian
 with short ranged (on-site) interactions, with the strength chosen
 so as to fit the vertical 
absorption peak in polyacetylene thin films. Furthermore, our calculations,
 being for polyenes, necessarily use
open boundary conditions.  For an even site chain there is one more short
bond than there are long bonds.
 This means that the ground state is
non-degenerate, as it is energetically unfavourable to swap long
 and short bonds, and is one reason for the long range confinement.
 The r\^{o}le played by boundary conditions is
 subtle and important, as real systems, such
 as oligomers and polymers with disorder, have a finite conjugation length.
 
Another consideration is the adequacy of the 
soliton fits used in generating the
 adiabatic energy
 curves. We consider the generalised potential energy curves
 where, for a given $x_0$, we allow $\xi$ to vary so as to minimise the
 energy. Results for the $1^1B_u^-$ state are included in
 Fig.\ \ref{potential_energy_curves}, which show that relaxing $\xi$ yields
 a substantial reduction in the energy, implying weak soliton/anti-soliton
 binding. However, the energy reduction 
for the $1^3B_u^+$ state is insignificant over the range of $x_0$ values
 plotted, indicating that there is a stronger binding of the solitons
 in the $2^1A_g^+$ and $1^3B_u^+$ states.
A further generalisation of the soliton fits would be to consider multiple
 soliton/anti-soliton pairs.

Finally, we note the consequences of our results for the interpretation
 of experiments. Our results for small polyenes are in good agreement with 
experiment---the energy difference of ca.\ 0.3 eV for the $1^1B_u^-$ state can 
probably be explained by solvation effects \cite{yaron97}, supporting
 the notion that 
the covalent $2^1A_g^+$ state is less polarised than the ionic
 $1^1B_u^-$. In the 
bulk limit the $1^1B_u^-$ and $2^1A_g^+$ energies are ca.\ 0.8 eV higher
 than data 
from linear \cite{vardeny} and 2-photon \cite{halvorson93} absorption and third
harmonic generation \cite{fann89} experiments on PA thin films. This implies
 that there are more 
substantial energy decreases due to solvation and aggregation (interchain
 hopping and 
eximer formation) effects. Such effects must be investigated via coupled
 chain 
calculations. Also, the neglect of quantum fluctuations in
 the adiabatic treatment of the lattice \cite{mckenzie},
leading to the pinning of the soliton structures, will  contribute
 to this energy difference. A full treatment must include dynamical
 phonons. Such a treatment would also increase
 our understanding of the soliton confinement.

R. J. B. was supported by the Australian Research Council.  
W. B. gratefully acknowledges financial support from the
Royal Society and the Gordon Godfrey Bequest of the UNSW.
We thank E. Jeckelmann and Y. Shimoi for useful discussions.
The calculations were performed at the
New South Wales Center for Parallel Computing.

\begin{table}[h]
\caption{
Convergence of the ground state 
%($1^1A_g^+$)
energy  (in eV) as a function of the 
%superblock Hilbert space size 
SBHSS for the $N = 102$ site system in the non-interacting case ($ U = 0 $) for three 
geometries defined by the soliton form (\protect\ref{2_soliton_form}), taking $\xi = 
4.03$ and (i) $x_0 = 0$ (uniformly dimerised geometry), (ii) $x_0 = \infty$ 
(uniformly dimerised with long and short bonds reversed), and (iii) $x_0 = 24.87$ (a 
geometry with a kink/anti-kink pair placed 1/4 and 3/4 of the way along the lattice). 
$m$ is the number of states retained per block.
}
\begin{tabular}{ccccc}
$m$ & SBHSS & $x_0 = 0$ & $x_0 = \infty$ & $x_0 = 24.87$ \\
\hline
75    & 5920  & $-$332.57146 & $-$330.205 & $-$331.333 \\
100   & 9384  & $-$332.57217 & $-$330.403 & $-$331.366 \\
150   & 22392 & $-$332.57257 & $-$330.426 & $-$331.407 \\
200   & 37512 & $-$332.57271 & $-$330.439 & $-$331.422 \\
230   & 52312 & $-$332.57272 & $-$330.446 & $-$331.428 \\
270   & 72392 & $-$332.57273 & $-$330.448 & $-$331.430 \\
EXACT &  ---  & $-$332.57276 & $-$330.452 & $-$331.434 \\
\end{tabular}
\label{U=0_convergence}
\end{table}

\begin{table}[h]
\caption{
Convergence of the ground state ($1^1A_g^+$) energy  and vertical and 0-0 transition 
energies of the 
%two-photon 
$2^1A_g^+$  and 
%one-photon 
$1^1B_u^-$ states as a function of the 
%superblock Hilbert space size
SBHSS for the $N = 102$ site system.
%and triplet ($1^3B_u^+$)
}
\begin{tabular}{cccccccc}
SBHSS & $1^1A_g^+$ &
% $E^{\text v}(2^1A_g^+)$ & $E^{\text{0-0}}(2^1A_g^+)$
 $2^1A_g^+$ & $2^1A_g^+$(0-0)
% & $E^{\text v}(1^1B_u^-)$ & $E^{\text{0-0}}(1^1B_u^-)$ \\
& $1^1B_u^-$ & $1^1B_u^-$(0-0) \\
%& $E^{\text v}(1^3B_u^+)$ & $E^{\text{0-0}}(1^3B_u^+)$ \\
%
\hline
15844 & $-$509.6330807 & 2.8927 & 1.8051 & 2.7719 & 2.6785 \\
%& 1.6039 & 1.1485 \\
25492 & $-$509.6330971 & 2.8795 & 1.8008 & 2.7650 & 2.6483 \\
%& 1.6005 & 1.1385 \\
36312 & $-$509.6331002 & 2.8764 & 1.7972 & 2.7617 & 2.6392 \\
%& 1.5996 & 1.1366 \\
54916 & $-$509.6331009 & 2.8744 & 1.7963 & 2.7605 & 2.6345 \\
%& 1.5990 & 1.1354 \\
67240 & $-$509.6331010 & 2.8737 & 1.7959 & 2.7601 & 2.6336 \\
%& 1.5989 & 1.1350 \\
%
\end{tabular}
\label{DMRG_convergence}
\end{table}

%\begin{table}[h]
%
%\caption{
%
%Fitting parameters for soliton form fits (\protect\ref{2_soliton_form}) and 
%(\protect\ref{4_soliton_form}) of the relaxed geometries of various excited states in 
%the $N = 102$ site polyene. 
%
%}
%
%\begin{tabular}{cccc}
%
%State      &   $\xi$  &  $x_0$  &  $x_d$  \\
%$1^3B_u^+$ &  4.01(3) & 4.96(2) & ---     \\
%$1^1B_u^-$ & 12.12(3) & 5.15(2) & ---     \\
%$E_g$      & 10.46(3) & 4.78(2) & ---     \\
%$2^1A_g^+$ &  4.93(2) & 4.02(1) & 5.41(2) \\
%
%\end{tabular}
%
%\label{fitting_parameters}
%
%\end{table}

\begin{references}
\bibitem{comment0}
We have checked the DMRG convergence of the $\delta_i$ and found that, for the $N=102$ site lattice, they are resolved to within 1\% or less.
\end{references}
\end{document}